\documentclass[a4paper,conference]{IEEEtran}
\usepackage{amsmath,amsfonts,graphicx,booktabs}
\usepackage{balance}

\DeclareMathOperator*{\argmax}{arg\,max}
\DeclareMathOperator*{\minimize}{minimize}
\DeclareMathOperator*{\subjectto}{subject\,to}

\ifCLASSINFOpdf
\else
\fi
\ifCLASSOPTIONcompsoc
  \usepackage[caption=false,font=normalsize,labelfont=sf,textfont=sf]{subfig}
\else
  \usepackage[caption=false,font=footnotesize]{subfig}
\fi
\begin{document}

%
\title{Mitigating Observation Biases in \\ Crowdsourced Label Aggregation}

\author{\IEEEauthorblockN{Ryosuke Ueda}
\IEEEauthorblockA{Kyoto University\\
Kyoto, Japan\\
rueda@ml.ist.i.kyoto-u.ac.jp}
\and
\IEEEauthorblockN{Koh Takeuchi}
\IEEEauthorblockA{Kyoto University\\
Kyoto, Japan\\
takeuchi@i.kyoto-u.ac.jp}
\and
\IEEEauthorblockN{Hisashi Kashima}
\IEEEauthorblockA{Kyoto University\\
Kyoto, Japan\\
kashima@i.kyoto-u.ac.jp}}


%


\maketitle

\renewcommand{\thefootnote}{\fnsymbol{footnote}}
\footnote[0]{© 2022 IEEE. Personal use of this material is permitted. Permission from IEEE must be 
obtained for all other uses, in any current or future media, including 
reprinting/republishing this material for advertising or promotional purposes, creating new 
collective works, for resale or redistribution to servers or lists, or reuse of any copyrighted 
component of this work in other works.}
\renewcommand{\thefootnote}{\arabic{footnote}}

\begin{abstract}
Crowdsourcing has been widely used to efficiently obtain labeled datasets for supervised learning from large numbers of human resources at low cost. 
However, one of the technical challenges in obtaining high-quality results from crowdsourcing is dealing with the variability and bias caused by the fact that it is
humans execute the work, and various studies have addressed this issue to improve the quality by integrating redundantly collected responses.
In this study, we focus on the observation bias in crowdsourcing. Variations in the frequency of worker responses and the complexity of tasks occur, which may affect the aggregation results when they are correlated with the quality of the responses. 
We also propose statistical aggregation methods for crowdsourcing responses that are combined with an observational data bias removal method used in causal inference.
Through experiments using both synthetic and real datasets with/without artificially injected spam and colluding workers, we verify that the proposed method improves the aggregation accuracy in the presence of strong observation biases and robustness to both spam and colluding workers.
\end{abstract}


%
\IEEEpeerreviewmaketitle

\section{Introduction}
Owing to the rapid development of machine learning technologies, there has been growing demand for data driven, human decision support in various fields.
In particular, prediction through supervised learning is one of the key techniques; however, its use requires accurate labels to train the prediction machine.
However, because these labels often need to be provided by humans, data collection can be extremely expensive.
The use of crowdsourcing is quite effective in obtaining a large number of training data, for example, in entity resolution~\cite{Liu2012-iw}, text classification~\cite{Wang2012-qb}, and image recognition~\cite{Su2012-zu}.
This is because crowdsourcing platforms such as Amazon Mechanical Turk can make large amounts of human resources available at a relatively low cost on the Internet.

One of the major challenges faced in training data collection using crowdsourcing is the variability in the reliability of the responses provided by the workers~\cite{Li2016-ux}.
This is attributed to the significant variations in the abilities and motivations of the human workers, as well as the anonymity of crowdsourcing workers.
In addition, there are spam workers who provide random answers without looking at the tasks~\cite{Raykar2012-is,Kittur2008-la}, as well as colluding workers who share their answers with other workers~\cite{Difallah2012-de,Chen2018-gt,khudabukhsh2014-kk}, which also contribute to the variability in the reliability.
To reduce the impact of incorrect responses, several studies have attempted to improve the quality by aggregating multiple redundantly collected responses from different workers.
This can be achieved using simple majority voting or statistical response aggregation methods that consider worker ability and the task difficulty~\cite{Dawid1979,Whitehill2009}.

In crowdsourcing, each worker does not necessarily need to answer all tasks in general.
Depending on the knowledge and preferences of the workers, or the type and difficulty of the tasks, the questions they answer may be biased.
In fact, we examined several public datasets, and found a large variation in worker response frequency and the biased relationship between such frequency and worker ability (Figure~\ref{fig:real}, Table~\ref{tab:corr}).
In the aforementioned response aggregation methods, workers with a high response frequency have a large influence on the aggregation results; 
therefore, the aggregation results can also be biased toward such high-frequency workers.
We investigate the effects of such biases on the performance of the aggregation methods, and attempt to improve the aggregation by removing bias when it has a negative influence.
We propose methods combining inverse propensity scoring~\cite{Rosenbaum1983-ij,Imbens2015}, which is an observation bias removal method used in causal inference, with simple majority voting, Dawid-and-Skene model (D\&S)~\cite{Dawid1979}, and 
GLAD (Generative model of Labels, Abilities, and Difficulties)~\cite{Whitehill2009}.
Our method estimates the integration results for uniformly random observations from biased observation labels.

Experiments using synthetic data suggest that bias removal improves the aggregation accuracy when there is a negative correlation between the observation rates of  the worker responses and agreement rates with true labels; however, the opposite occurs when there is a positive correlation.
In addition, an examination of some real datasets indicates weak negative correlations between the number of worker responses and the percentage of correct responses. For datasets with such a negative correlation, the proposed method outperforms the baseline when the number of labels is small, and thus the effect of the observation bias increases.
Because an analysis of real datasets suggests the existence of spam workers who provide a significant number of random answers, we created semi-synthetic datasets with enhanced observation biases by hypothetically highlighting such inappropriate workers. 
The results of the additional experiments indicate that the proposed methods are robust to the presence of both spam and malicious colluding workers.

Our contributions are summarized as follows:
\begin{itemize}
    \item We investigate and control the effect of observation bias on the results of crowdsourced label aggregation.
    \item We propose an EM algorithm-based method for label aggregation using a novel lower bound that mitigates the observation bias.
\end{itemize}

\section{Related Work}
Quality control is one of the major challenges of crowdsourced label aggregation, and how to handle the uncertainty brought by workers is an important aspect of the problem.
For example, the Dawid-and-Skene model~\cite{Dawid1979} models a worker as a confusion matrix,
where an EM algorithm simultaneously estimates the confusion matrices and the ground truth instance labels.
GLAD~\cite{Whitehill2009} considers instance difficulty as well as worker ability based on item response theory, and uses an EM algorithm to estimate the parameters and ground truth labels.
Learning from crowds (LFC)~\cite{Raykar2012-is} is a problem setting that directly learns a classifier from task instance features and crowdsourced labels.
The ground truth estimates are also obtained as byproducts.
LFC can be considered as an extension of D\&S to cases where task instance features are available.
Additionally, various label aggregation algorithms based on Bayesian inference have been proposed that aim to deal with small number of worker labels.
Bayesian Classifier Combination (BCC)~\cite{Kim2012-wh} is a Bayesian extension of D\&S and uses MCMC for inference. Community BCC~\cite{Venanzi2014-fx} further extends BCC to consider group structures within workers.
In recent years, Enhanced BCC~\cite{Li2019-EBCC}, which models correlation between workers, shows better performance on many datasets.
In addition, Li et al.~\cite{Li2019-BWA} proposes a Bayesian model without worker's confusion matrix.
Our proposal in the present study is to investigate and control the effects of observational biases in crowdsourced label aggregation.
In order to test the promise of the idea, we restrict our focus to basic aggregation methods such as majority voting, D\&S, and GLAD, as our base model in the present study.
Although more modern Bayesian models like BCC are the state-of-the-art methods, implementing the proposed idea on them is not necessarily obvious, and is a subject for future research.

Recent work explores the effect of biases in crowdsourced label aggregation induced by crowdworkers being humans.
Eickhoff~\cite{Eickhoff2018-eb} investigates the impact of several cognitive biases such as the bandwagon effect in crowdsourced experiments, and shows that inappropriate task design leads to poor accuracy.
Zhuang and Young~\cite{Zhuang2015-ik} show that when multiple tasks are annotated by a worker as a single batch, the combination of tasks in a batch can affect the response.
In addition, both crowdsourcing~\cite{Newell2016-ko} and psychological~\cite{Huang2018-ge} experiments showed that the effect of previous tasks is present when performing sequential tasks.
Biswas et al.~\cite{Biswas2020-me} focus on worker race in a defendant recidivism prediction task, reporting that classifiers are fair when trained with balanced worker racial distribution data.
In addition, the existence of a ``confirmation bias" that is responsive to choices that fit workers' beliefs~\cite{Coscia2020-fr,La_Barbera2020-hr,Hube2019-nf} is shown.
In contrast, we focus on the observation bias and investigate and control its effects in this study.

In crowdsourcing, it is desirable to avoid spammers who respond randomly to many instances for reward, and the presence of spammers reinforces the observational bias.
Kittur et al.~\cite{Kittur2008-la} show the effectiveness of introducing a rating task at the beginning of the entire task.
Raykar and Yu~\cite{Raykar2012-is} propose scores for spammer detection.

Several studies explore issues related to observation bias.
Han et al.~\cite{Han2019-zl} investigate ``task abandonment," which was once tackled by a worker but never submitted.
Difallah et al.~\cite{Difallah2012-de} use the propensity score to estimate the number of workers on Amazon Mechanical Turk.
Schnabel et al.~\cite{Schnabel2016-xa} mitigate observation bias in recommendation systems using weighting by the inverse of the propensity score.
\section{Problem Setting}
In this study, we consider a standard problem setting for crowdsourced label aggregation.
Suppose we have $m$ task instances such as a set of images and texts. 
Each task instance belongs to one of $K$ different classes $\{1,\ldots,K\}$; 
the ground truth label of the $j$-th instance is denoted by $Z_j \in \{1,\ldots,K\}$.
We assume that the set of ground truth labels for the task instances $Z_1, Z_2, \ldots, Z_m$ are unknown.

We ask crowdsourcing worker $i \in \{1, \ldots, n\}$ to give labels to $m$ task instances
denoted by $L_{ij} \in \{1,\ldots,K\}$, which is the label given by worker $i$ to task instance $j$. 
The workers do not necessarily have to give labels to all task instances, i.e., $L_{ij}$ is occasionally missing.
We introduce $O_{ij}$ as a variable to indicate whether the label $L_{ij}$ is obtained; we set $O_{ij}=1$ when $L_{ij}$ is observed; otherwise, we set $O_{ij}=0$.

Our goal is to estimate the ground truth labels $Z = \{ Z_1, Z_2, \ldots, Z_m \}$ by aggregating the set of crowdsourced labels $L = \{L_{ij}: O_{ij}=1\}$.
Although the simplest way to aggregate the crowdsourced labels is majority voting, in recent years more sophisticated probabilistic generative models have been used.
A number of models have also been proposed for considering various factors such as worker ability and task difficulty, and for estimating these parameters along with the ground truth answers~\cite{Quoc2013, Sheshadri2013-cb}.
However, such models do not account for observation biases, and workers who respond more frequently have a larger impact on the results. The accuracy may also decrease when there are biases in the response frequency and reliability.

\section{Proposed Method}
We propose the use of response aggregation methods to reduce the effect of observation bias discussed in the previous section.
We combine inverse propensity scoring (IPS)~\cite{Rosenbaum1983-ij,Imbens2015}, which is used to remove observation bias in causal inference, with three aggregation methods:
simple majority voting, D\&S and GLAD.
\subsection{Majority-Voting-based Method: IPS-MV}
First, we propose IPS-MV, which combines the simple majority voting (MV) method with IPS.
With the simple MV method, we obtain the aggregate label $\hat{Z}_j$ for task instance $j$ as
\begin{align}
    \hat{Z}_j = \argmax_k a_j^{(k)},~ 
    a_j^{(k)} = \sum_{(i,j):O_{ij}=1} I(L_{ij} = k), \label{eq:MV}
\end{align}
where $I$ represents the indicator function.

When we adopt IPS, instead of the equally weighted aggregation formula~\eqref{eq:MV}, we apply its weighted version: 
\begin{align*}
    {a^\text{IPS}}_j^{(k)} = \sum_{(i,j): O_{ij}=1} \frac{1}{e_{ij}}  I(L_{ij} = k),
\end{align*}
where 
$
    e_{ij} = \Pr[ O_{ij}=1 ]   
$
is the (estimated) probability that worker $i$ answers instance $j$, which is called the \emph{propensity score}.

\subsection{EM-Algorithm-based Method: IPS-D\&S and IPS-GLAD}
Next, we propose IPS-D\&S and IPS-GLAD, which combine IPS with D\&S and GLAD, well-known label aggregation methods using the EM algorithm, in order to mitigate observation bias.
D\&S~\cite{Dawid1979} is one of the early representative approaches to this problem, and shows high accuracy in decision-making tasks~\cite{Zheng2017-tp}.
GLAD~\cite{Whitehill2009} is an algorithm that extends D\&S to simultaneously handle worker ability and task difficulty.
Although D\&S and GLAD have different label generation assumptions, they can be estimated with the same EM algorithm framework.

\begin{figure}[tb]
    \centering
    \fbox{\subfloat[D\&S\label{fig:ds-gm}]{\includegraphics[bb=130 150 410 330, width=0.45\linewidth, clip=true]{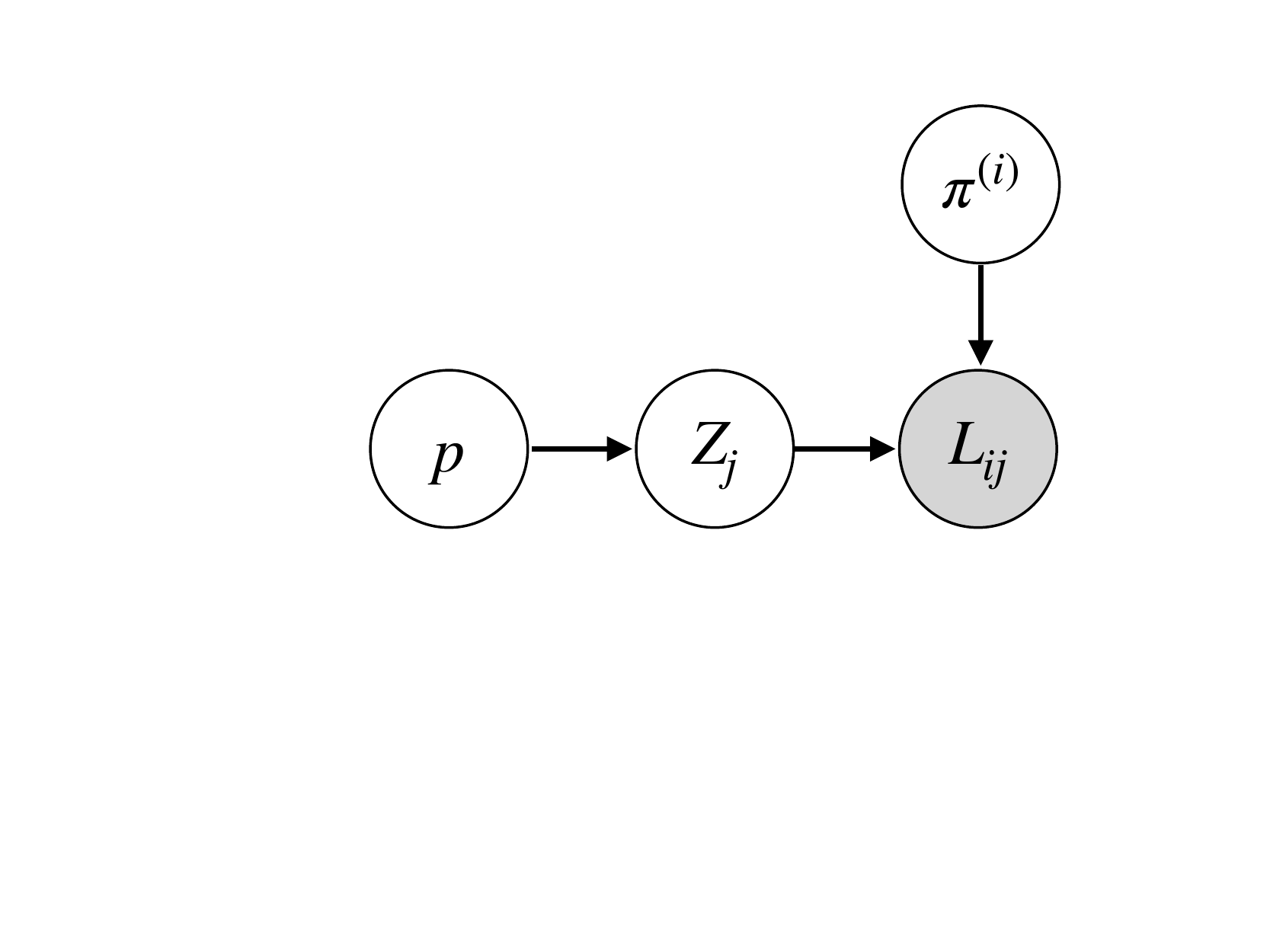}}}
    \fbox{\subfloat[GLAD\label{fig:glad-gm}]{\includegraphics[bb=130 150 410 330, width=0.45\linewidth, clip=true]{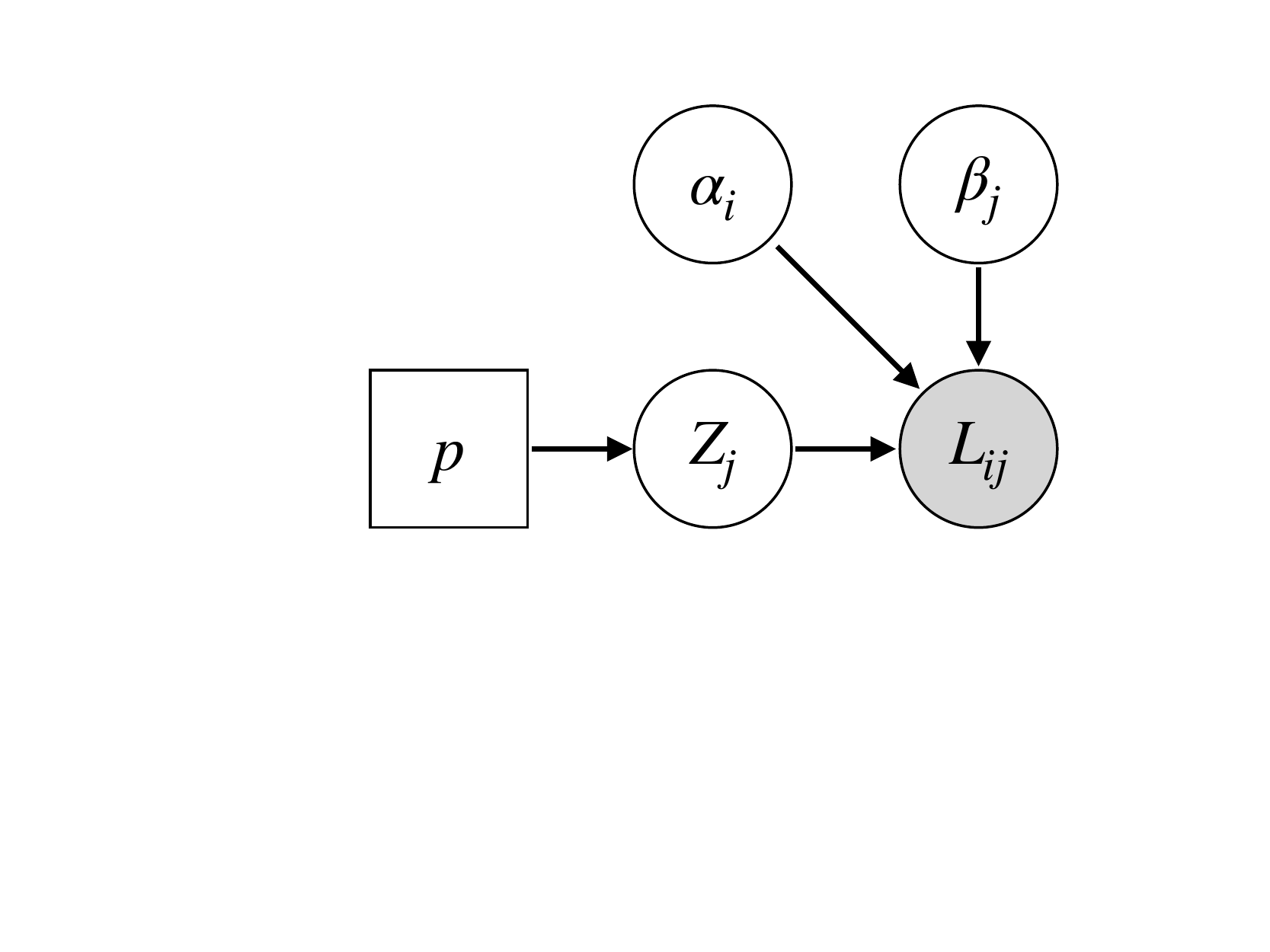}}}
    \caption{Graphical model of the label generation process. Only shaded variables are observed.}
    \label{fig:gm}
\end{figure}

Figure~\ref{fig:ds-gm} shows the graphical model of D\&S; 
$\pi^{(i)}_{k k'}$ indicates the conditional probability that worker $i$ will respond label $k'$ given true label $k$.
Hence, the likelihood of the label is given as 
\[
p(L_{ij} = k' \mid Z_j = k, \pi^{(i)}_{k k'}) = \pi^{(i)}_{k k'}.
\]
In D\&S, we optimize the label prior $p = p(Z)$ as a parameter.

The graphical model of GLAD is shown in Figure~\ref{fig:glad-gm}.
In the GLAD model, the probability that worker $i$ provides the correct answer $Z_j$ to instance $j$ depends on the worker's ability $\alpha_i \in (-\infty, +\infty)$, and task difficulty $1/\beta_j \in (0, +\infty)$, which are specifically defined as 
\[
    p(L_{ij} = k \mid Z_j = k, \alpha_i, \beta_j) = \sigma(\alpha_i \beta_j),
\]
where $\sigma$ denotes the logistic function.
When the answer $L_{ij}$ is incorrect, it is sampled from a uniform distribution over $K-1$ incorrect labels.
The probability that a wrong answer $k' \neq k$ is given is 
\[
    p(L_{ij} = k' \mid Z_j = k, \alpha_i, \beta_j) = \frac{1 - \sigma(\alpha_i \beta_j)}{K - 1}.
\]
We set the label prior $p$ to a uniform distribution in this study.

Let $\boldsymbol{\theta}$ be the unobserved parameters other than $Z$ and $p$ (i.e. $\{\pi^{(i)}\}_{i=1}^n$ in D\&S, and $\{\alpha_i\}_{i=1}^n, \{\beta_j\}_{j=1}^m$ in GLAD).
The maximum likelihood estimation is used to estimate the distribution $q(Z_j)$ of the true label $Z_j$, the model parameter $\boldsymbol{\theta}$, and (for D\&S only) the prior distribution $p$.
Instead of maximizing the log likelihood $\ln p(L \mid \boldsymbol{\theta})$, its lower bound:
\begin{equation*}
\begin{split}
    \mathcal{L}(q, \boldsymbol{\theta}) 
    &=\sum_{(i,j): O_{ij}=1} \sum_{Z_j} q(Z_j) \ln p(L_{ij} \mid Z_j, \boldsymbol{\theta})
    \\&\quad+ \sum_{j,Z_j} q(Z_j) \ln \frac{p(Z_j)}{q(Z_j)},
\end{split}\label{eq:lb}
\end{equation*}
is maximized using the EM algorithm.

The lower bound $\mathcal{L}$ is rewritten as
\begin{equation*}
\begin{split}
    \mathcal{L}(q, \boldsymbol{\theta}) 
    &=\sum_{i,j} \sum_{Z_j} O_{ij} q(Z_j) \ln p(L_{ij} \mid Z_j, \boldsymbol{\theta})
    \\&\quad+ \sum_{j,Z_j} q(Z_j) \ln \frac{p(Z_j)}{q(Z_j)}.
\end{split}
\end{equation*}
This indicates that the propensity score $\Pr[O_{ij}=1]$ implicitly weights the lower bound $\mathcal{L}$, 
and the lower bound is biased when the propensity scores are biased.
Instead, we use the following unbiased lower bound $\mathcal{L}^{\text{IPS}}$:
\begin{equation}
\begin{split}
    \mathcal{L}^{\text{IPS}}(q, \boldsymbol{\theta}) 
    &=\sum_{(i,j): O_{ij}=1} \sum_{Z_j} \frac{q(Z_j)}{e_{ij}} \ln p(L_{ij} \mid Z_j, \boldsymbol{\theta})
    \\&\quad+ \sum_{j,Z_j} q(Z_j) \ln \frac{p(Z_j)}{q(Z_j)}.
\end{split}\label{eq:ipslb}
\end{equation}
Because the expected value for $\mathcal{L}^{\text{IPS}}$ over $O$ is
\begin{equation*}
\begin{split}
    \mathbb{E}_O[\mathcal{L}^{\text{IPS}}] &=
    \sum_{i,j} \sum_{Z_j} q(Z_j) \ln p(L_{ij} \mid Z_j, \boldsymbol{\theta})
    \\&\quad+ \sum_{j,Z_j} q(Z_j) \ln \frac{p(Z_j)}{q(Z_j)},
\end{split}
\end{equation*}
the new lower bound (\ref{eq:ipslb}) is unbiased w.r.t. the uniform distribution. The new lower bound can also be maximized using the EM algorithm.

\subsection{Propensity Score Estimation}
Because true propensity scores are not always available in practice, 
we estimate them using a 1-bit matrix completion (1-bit MC)~\cite{Davenport2014}, which applies a matrix completion under a nuclear-norm ($\| \cdot \|_*$) constraint.
The 1-bit MC approximates $O_{ij}$ as $\sigma(A_{ij})$ using a  matrix $A \in \mathbb{R}^{n \times m}$.
The optimization problem w.r.t. $A$ is given as
\begin{align*}
    & \minimize_{A \in \mathbb{R}^{n \times m}} \quad -\sum_{i, j} O_{ij} \ln \sigma(A_{ij}) + (1 - O_{ij}) \ln (1 - \sigma(A_{ij}))\\
    & \subjectto \quad \| A \|_* \leq \gamma \sqrt{nm},
\end{align*}
where $\gamma > 0$ is a hyperparameter.
It is known that the nuclear norm is a convex relaxation of the rank constraint.
As $\gamma$ is decreased, $A$ approaches the low-rank matrix and eventually approaches a zero matrix; on the other hand, as $\gamma$ is increased, the constraint is relaxed and $\hat{e}_{ij}$ approaches $O_{ij}$.

We use the estimated $\sigma(A_{ij})$ as an approximation of the propensity score $\hat{e}_{ij}$.

\section{Experiments}
The proposed method estimates unbiased results by giving smaller weights to responses that are more likely to be observed. 
Depending on the relationship between the propensity and percentage of correct answers, the proposed model is expected to give different aggregation results from the base model. 
Therefore, we first investigate the relationship between the aggregation accuracy and correlations of propensity and the percentage of correct answers using synthetic datasets.
We further compare the results of different methods on real datasets.
Finally, we conduct experiments using semi-synthetic data with both virtual spam and colluding workers to investigate the robustness against such harmful workers.

\begin{figure}[tb]
    \centering
    \includegraphics[width=0.6\linewidth]{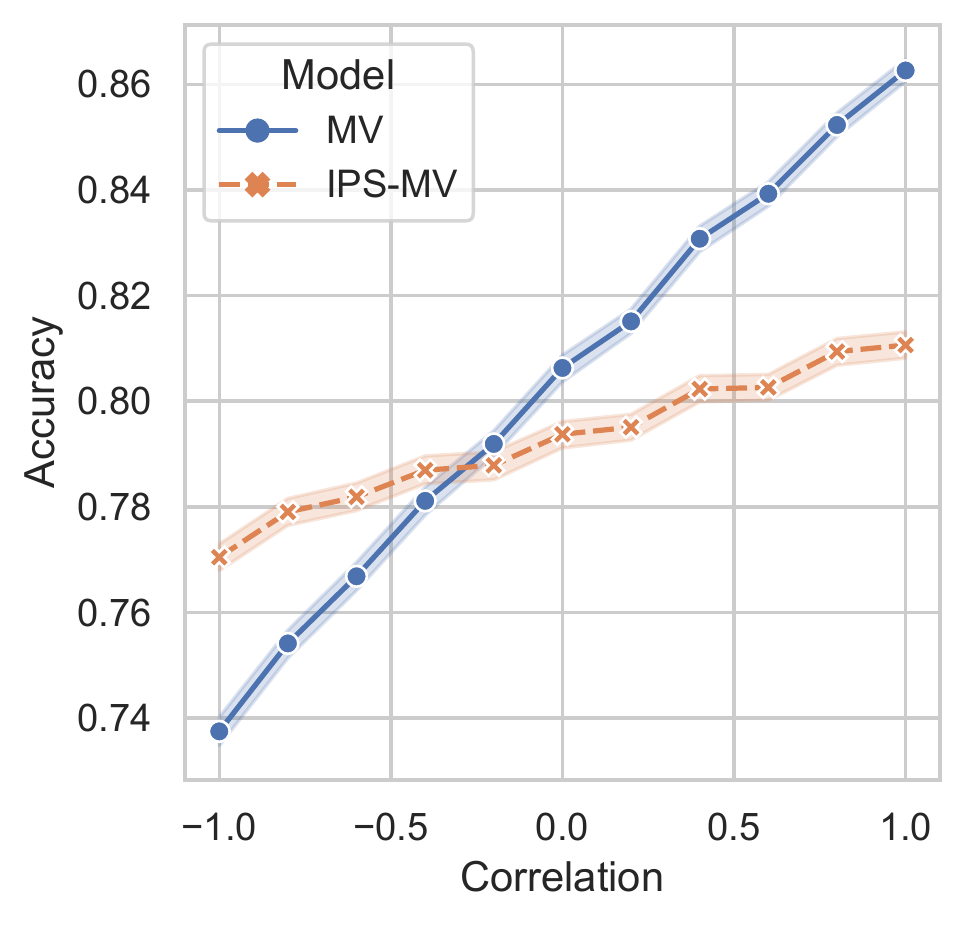}
    \caption{Performance comparison using synthetic datasets with different correlations between observation and correct answer probabilities.}
    \label{fig:synthetic}
\end{figure}

\begin{figure*}[tb]
    \centering
    \subfloat[RTE]{\includegraphics[width=0.235\linewidth]{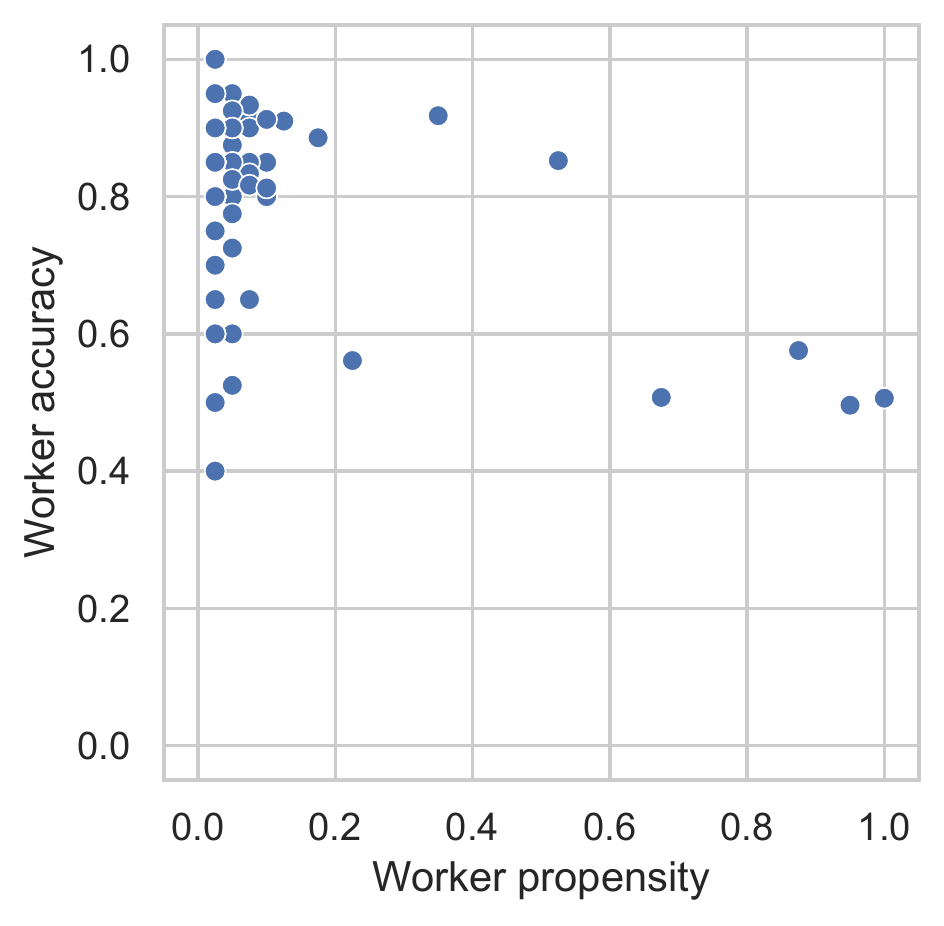}}\quad
    \subfloat[TEMP]{\includegraphics[width=0.235\linewidth, trim={0mm, 0mm, 0mm, 0mm}]{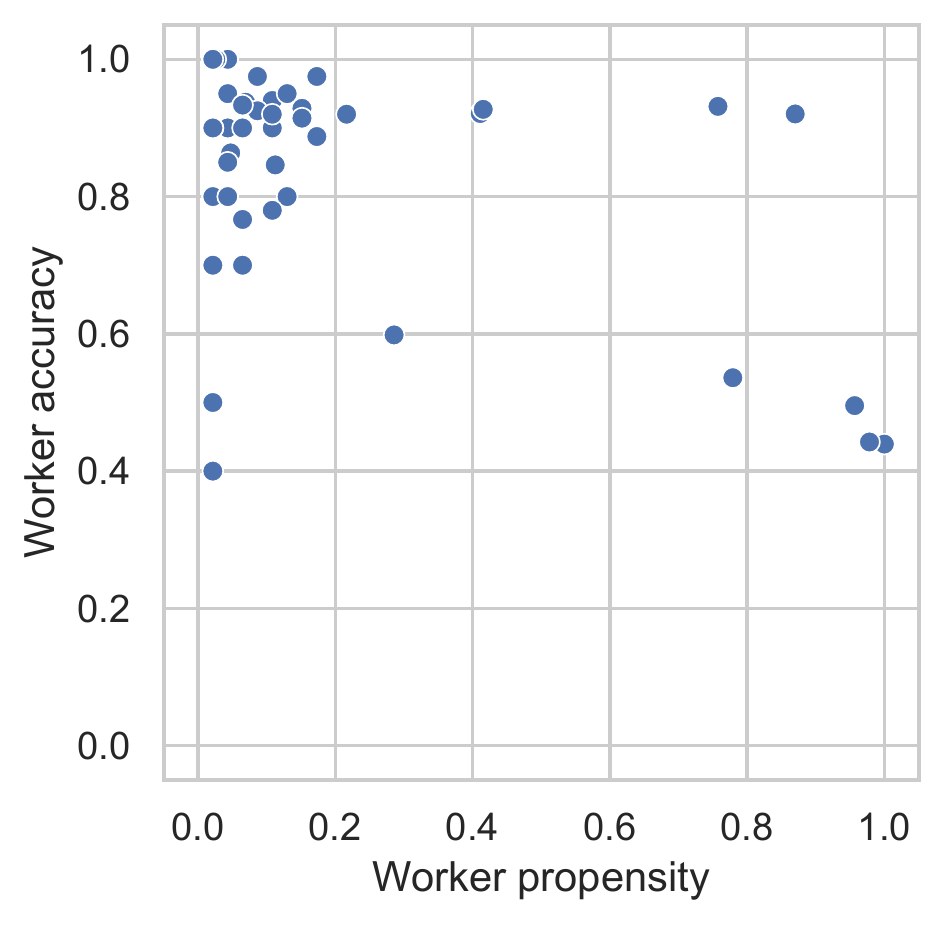}} \quad
    \subfloat[WSD]{\includegraphics[width=0.235\linewidth]{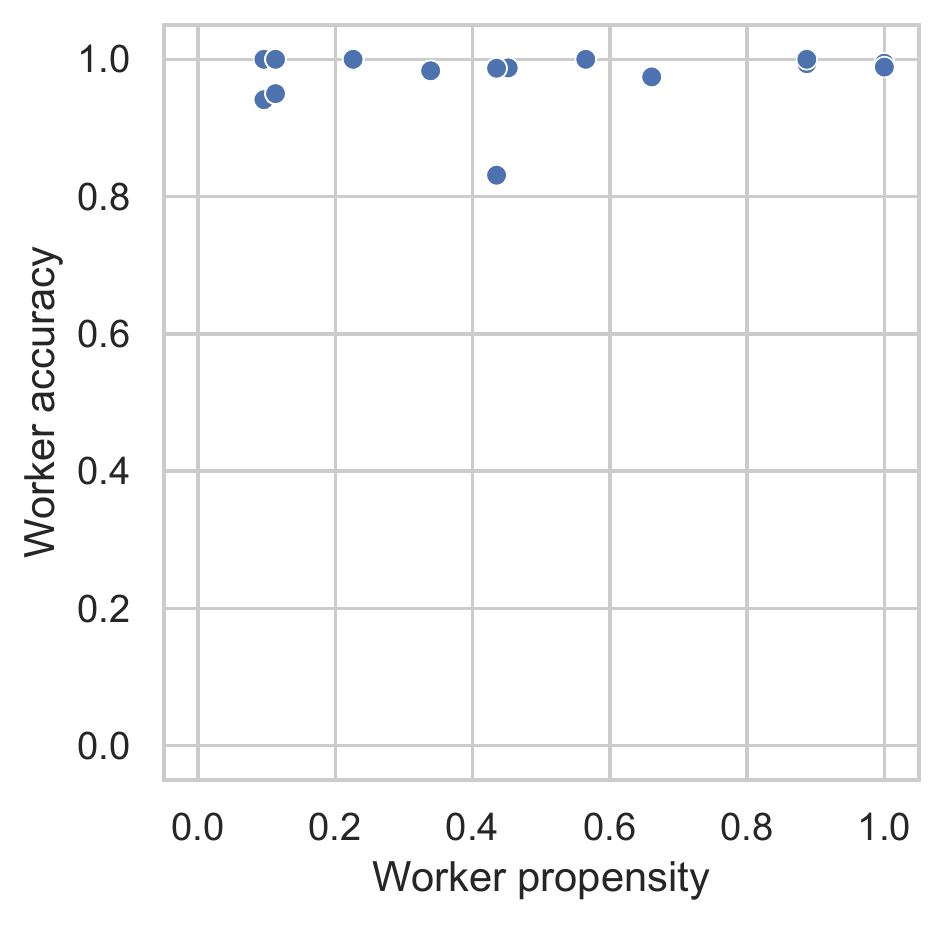}}\quad
    \subfloat[SP]{\includegraphics[width=0.235\linewidth]{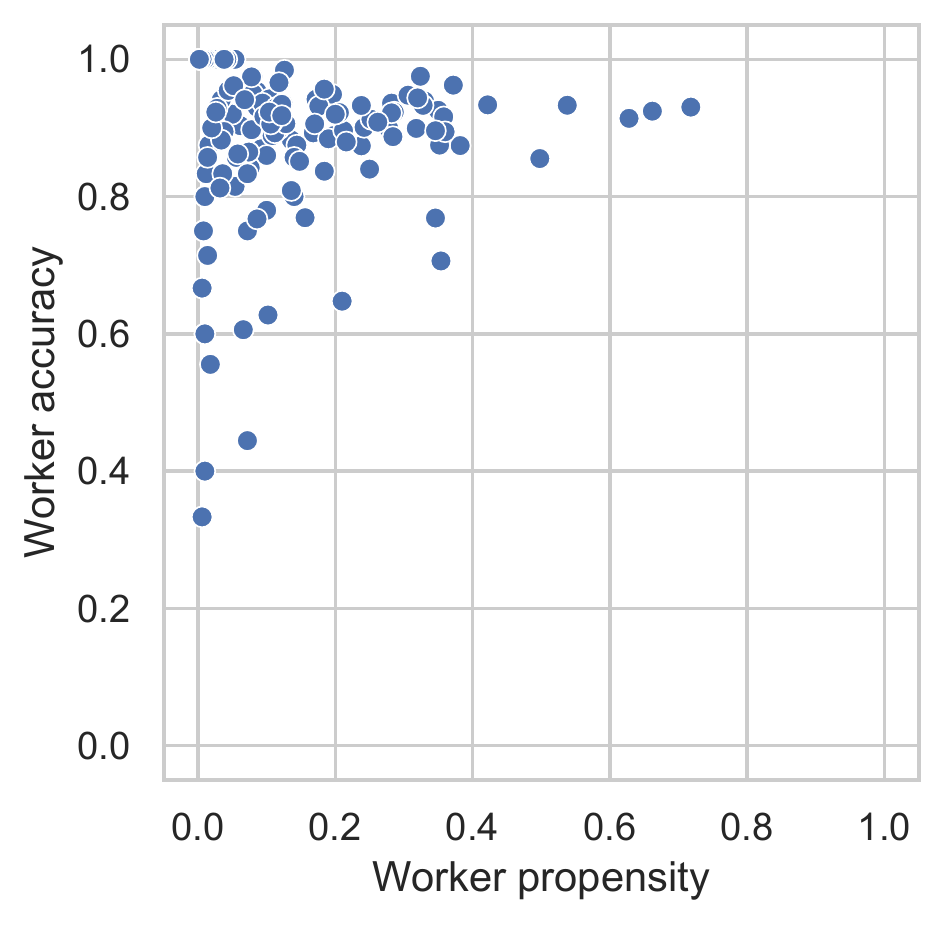}}
    \caption{Relationship between worker propensity and accuracy in the real datasets.}
    \label{fig:real}
\end{figure*}

\subsection{Experiments Using Synthetic Data}
First, we investigate the impact of the relationship between the propensity and accuracy on the aggregation performance of the proposed method using synthetic datasets.
A synthetic dataset includes 20 workers and 100 task instances.
The ground truth labels $Z$ are uniformly sampled at random over binary classes ($K=2$).
For each $(i,j)$-pair, we sample the observation probability $e_{ij}$ and the correct answer probability $c_{ij}$ from the two-dimensional Gaussian distribution with the mean $(0.15, 0.75)$ and variance $(0.075^2, 0.125^2)$ so that the average number of labels per instance is 3.
The covariances are determined according to the correlation coefficient between -1 and 1.
The sampled parameters are clipped to within $[0,1]$.

We compare the accuracy of the simple MV and the proposed bias-mitigating MV (IPS-MV) with the true observation rate $e_{ij}$.
The data generation and estimation are repeated 1000 times, and their average accuracy is compared.
In this experiment, we investigate the dependency of the correlation and accuracy, and do not consider worker ability. Hence, D\&S and GLAD are not used.
Figure \ref{fig:synthetic} shows the average accuracy when the correlation between the observation probability $e_{ij}$ and correct answer probability $c_{ij}$ is varied.
The stronger the negative correlation is, the better the proposed IPS-MV method in comparison to MV; by contrast, MV is more accurate for non-negative correlations.
The experimental results indicate that the relationship between the propensity and ability has a significant impact on the aggregation accuracy.
Note that since the correlation estimation requires ground-truth labels, it is usually impossible to determine in advance whether the proposed method will be effective.

\subsection{Experiments Using Real Data}
\begin{table}[tb]
    \centering
    {\tabcolsep=0.4mm
    \caption{Dataset Description.}
    \begin{tabular}{lrrrr}
        \toprule
        Dataset & \# classes & \# workers & \# instances & \# labels per instance \\
        \midrule
        (a)~RTE & 2 & 164 & 800 & 10 \\
        (b)~TEMP & 2 & 76 & 462 & 10 \\
        (c)~WSD & 3 & 34 & 177 & 10 \\
        (d)~SP & 2 & 143 & 500 & 20 \\
        \bottomrule
    \end{tabular}
    \label{tab:description}
    }
\end{table}

\begin{table}[tb]
    \centering
    {\tabcolsep=0.8mm
    \caption{Correlation Coefficient Between Worker Propensity and Accuracy in the Real Datasets.}
    \begin{tabular}{l|rrrr}
        \toprule
        Dataset & (a)~RTE & (b)~TEMP & (c)~WSD & (d)~SP \\
        \midrule
        Correlation & -0.384 & -0.377 & 0.062 & 0.097 \\
        \bottomrule
    \end{tabular}
    \label{tab:corr}
    }
\end{table}

The previous experiments on artificial data suggested that the removal of observation bias leads to an improvement when there is a negative correlation between the observation probability and the correct answer probability; therefore, we tested this hypothesis using the four real datasets:
(a) Recognizing Textual Entailment (RTE)~\cite{Snow2008},
(b) Temporal Ordering (TEMP)~\cite{Snow2008},
(c) Word Sense Disambiguation (WSD)~\cite{Snow2008}, and
(d) Sentiment Popularity\footnote{https://eprints.soton.ac.uk/376544/}.
Table~\ref{tab:description} shows the number of classes, workers, instances, and the number of labels per instance for each dataset.
Figure~\ref{fig:real} shows a plot of worker propensity versus accuracy, and Table~\ref{tab:corr} shows their correlation coefficients.
Among the four datasets, (c) WSD and (d) SP do not show much correlation, whereas (a) RTE and (b) TEMP show negative correlations, thus suggesting the occurrence of observation biases.
In addition, Figure~\ref{fig:real} shows the presence of spammers answering many tasks with correct answer rates around 50\% for binary questions (that is, the chance level) in the (a) RTE and (b) TEMP datasets.

\begin{table*}[t]
\centering
{\tabcolsep=2.0mm
\caption{Comparison of the Aggregation Accuracy by Different Methods.}
\begin{tabular}{lcccccccccccc}
\toprule
Dataset & \multicolumn{3}{c}{RTE} & \multicolumn{3}{c}{TEMP} & \multicolumn{3}{c}{WSD} & \multicolumn{3}{c}{SP} \\
\cmidrule(lr){2-4}\cmidrule(lr){5-7}\cmidrule(lr){8-10}\cmidrule(lr){11-13}
Number of labels per task &                     2 &                     5 &                     8 &                     2 &                     5 &                     8 &                     2 &                     5 &                     8 &                     2 &                     5 &                     8 \\
\midrule
MV                       &                 0.769 &                 0.845 &                 0.896 &                 0.789 &                 0.894 &                 0.939 &                 0.973 &                 0.992 &  $\boldsymbol{0.994}$ &                 0.882 &                 0.933 &                 0.938 \\
IPS-MV ($\gamma=0.1$)    &  $\boldsymbol{0.809}$ &                 0.845 &                 0.902 &                 0.825 &                 0.894 &                 0.939 &                 0.979 &  $\boldsymbol{0.993}$ &  $\boldsymbol{0.994}$ &                 0.880 &                 0.933 &                 0.937 \\
IPS-MV ($\gamma=1$)    &  $\boldsymbol{0.809}$ &                 0.867 &                 0.908 &                 0.825 &                 0.905 &                 0.937 &                 0.979 &                 0.992 &                 0.993 &                 0.880 &                 0.933 &                 0.938 \\
IPS-MV ($\gamma=10$)   &                 0.808 &                 0.871 &                 0.902 &                 0.824 &                 0.893 &                 0.933 &                 0.977 &                 0.992 &  $\boldsymbol{0.994}$ &                 0.880 &                 0.924 &                 0.928 \\
D\&S                       &                 0.757 &                 0.899 &                 0.925 &                 0.842 &  $\boldsymbol{0.929}$ &  $\boldsymbol{0.942}$ &                 0.988 &                 0.989 &                 0.993 &                 0.900 &  $\boldsymbol{0.938}$ &  $\boldsymbol{0.944}$ \\
IPS-D\&S ($\gamma=0.1$)    &                 0.767 &  $\boldsymbol{0.900}$ &  $\boldsymbol{0.927}$ &                 0.835 &  $\boldsymbol{0.929}$ &                 0.941 &                 0.984 &                 0.988 &                 0.991 &                 0.902 &                 0.937 &  $\boldsymbol{0.944}$ \\
IPS-D\&S ($\gamma=1$)    &                 0.781 &                 0.898 &                 0.926 &                 0.844 &                 0.926 &                 0.937 &                 0.980 &                 0.986 &                 0.989 &                 0.902 &                 0.935 &  $\boldsymbol{0.944}$ \\
IPS-D\&S ($\gamma=10$)   &                 0.798 &                 0.889 &                 0.922 &  $\boldsymbol{0.848}$ &                 0.925 &                 0.939 &                 0.988 &                 0.989 &                 0.993 &                 0.901 &                 0.928 &                 0.938 \\
GLAD                     &                 0.788 &                 0.894 &                 0.921 &                 0.835 &                 0.925 &                 0.940 &  $\boldsymbol{0.991}$ &  $\boldsymbol{0.993}$ &  $\boldsymbol{0.994}$ &  $\boldsymbol{0.904}$ &                 0.934 &  $\boldsymbol{0.944}$ \\
IPS-GLAD ($\gamma=0.1$)  &                 0.786 &                 0.895 &                 0.920 &                 0.836 &                 0.926 &                 0.939 &  $\boldsymbol{0.991}$ &  $\boldsymbol{0.993}$ &  $\boldsymbol{0.994}$ &  $\boldsymbol{0.904}$ &                 0.934 &  $\boldsymbol{0.944}$ \\
IPS-GLAD ($\gamma=1$)  &  $\boldsymbol{0.809}$ &                 0.890 &                 0.911 &                 0.846 &                 0.923 &                 0.935 &                 0.982 &  $\boldsymbol{0.993}$ &                 0.993 &                 0.900 &                 0.934 &                 0.941 \\
IPS-GLAD ($\gamma=10$) &  $\boldsymbol{0.809}$ &                 0.884 &                 0.910 &                 0.843 &                 0.921 &                 0.936 &                 0.988 &                 0.992 &  $\boldsymbol{0.994}$ &                 0.891 &                 0.924 &                 0.928 \\
\bottomrule
\end{tabular}
\label{tab:sampling2}
}
\end{table*}

Even though crowdsourcing is relatively low cost, we want to obtain high aggregation accuracy with as few labels as possible, but fewer labels are more sensitive to observation bias.
To investigate this situation, we conduct experiments using randomly sampled subsets of the original datasets.
The random sampling is conducted five times and the accuracy is averaged.
For propensity score estimation, we use the 1-bit MC.
Table~\ref{tab:sampling2} shows the results when the number of labels per task is set to 2, 5 and 8. For (a) RTE and (b) TEMP, which show a weakly negative correlation in Table~\ref{tab:corr}, the proposed method shows a higher accuracy than the baseline. The difference in accuracy is particularly large when the number of labels is small. For the two labels in RTE, IPS-MV outperforms MV by up to 4.0 percentage points, IPS-D\&S outperforms D\&S by up to 4.1 percentage points, and IPS-GLAD outperforms GLAD by up to 2.1 percentage points. By contrast, in (c) WSD and (d) SP, the accuracy of the proposed method is equal to or slightly lower than the baseline.

\subsection{Robustness against Harmful Workers}
The previous analyses of the real datasets suggest the existence of spammers in the RTE and TEMP datasets, which can be one of the factors causing observation biases.
To investigate the effect of such observation biases, we conducted experiments using semi-synthetic data with two types of harmful workers: spam workers and colluding workers.
The synthetic spam workers and colluding workers respond to all tasks. The spam worker labels are sampled from a uniform distribution over $\{1, \ldots, K\}$. 
The colluding workers are more malicious, and they collude and try to guide the outcome of the majority vote~\cite{Difallah2012-de,Chen2018-gt,khudabukhsh2014-kk}. In our experiment, a label is sampled from a uniform distribution over $\{1, \ldots, K\}$, and all the colluding workers respond with the same label.
We continue adding spam and collusion workers until the malicious worker labels make up 50\% of all labels.

\begin{figure*}
    \centering
    \subfloat[RTE\label{fig:spam-RTE}]{\includegraphics[width=0.45\linewidth]{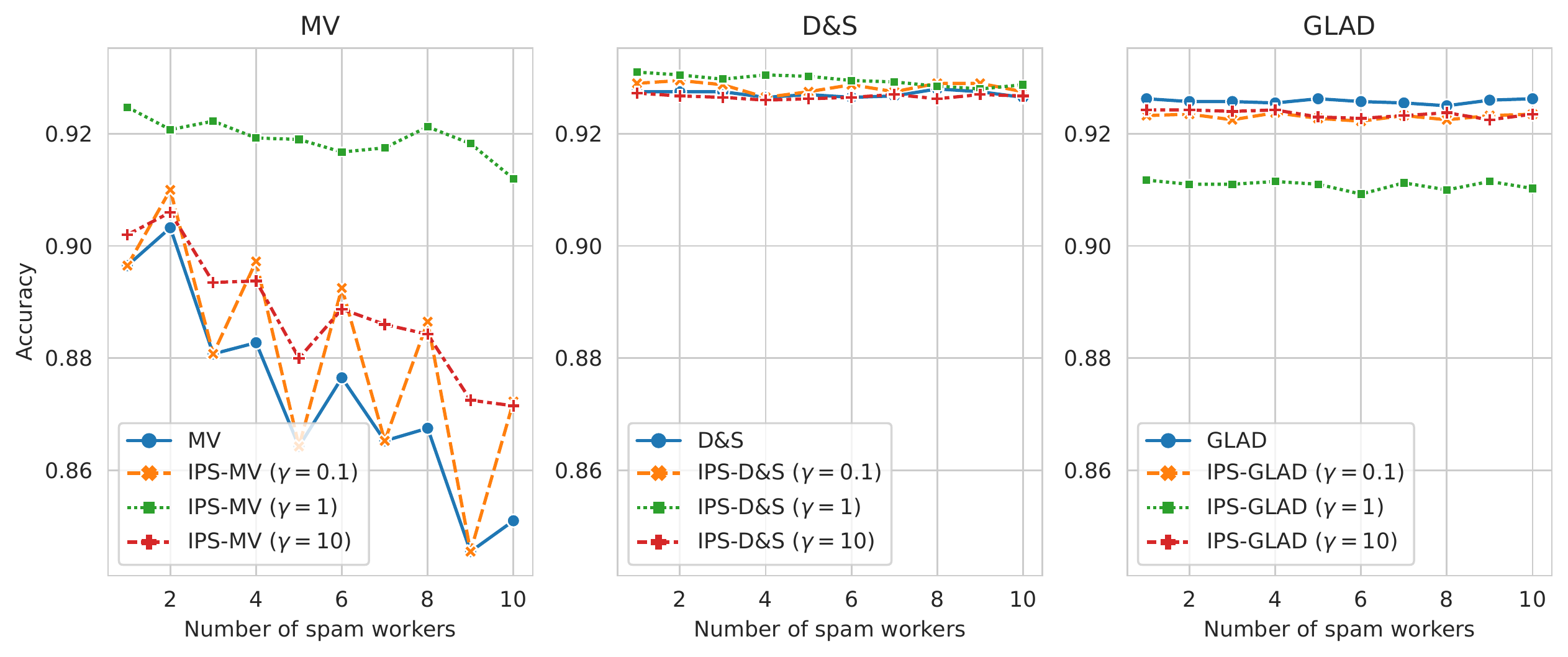}} \quad
    \subfloat[TEMP\label{fig:spam-TEMP}]{\includegraphics[width=0.45\linewidth]{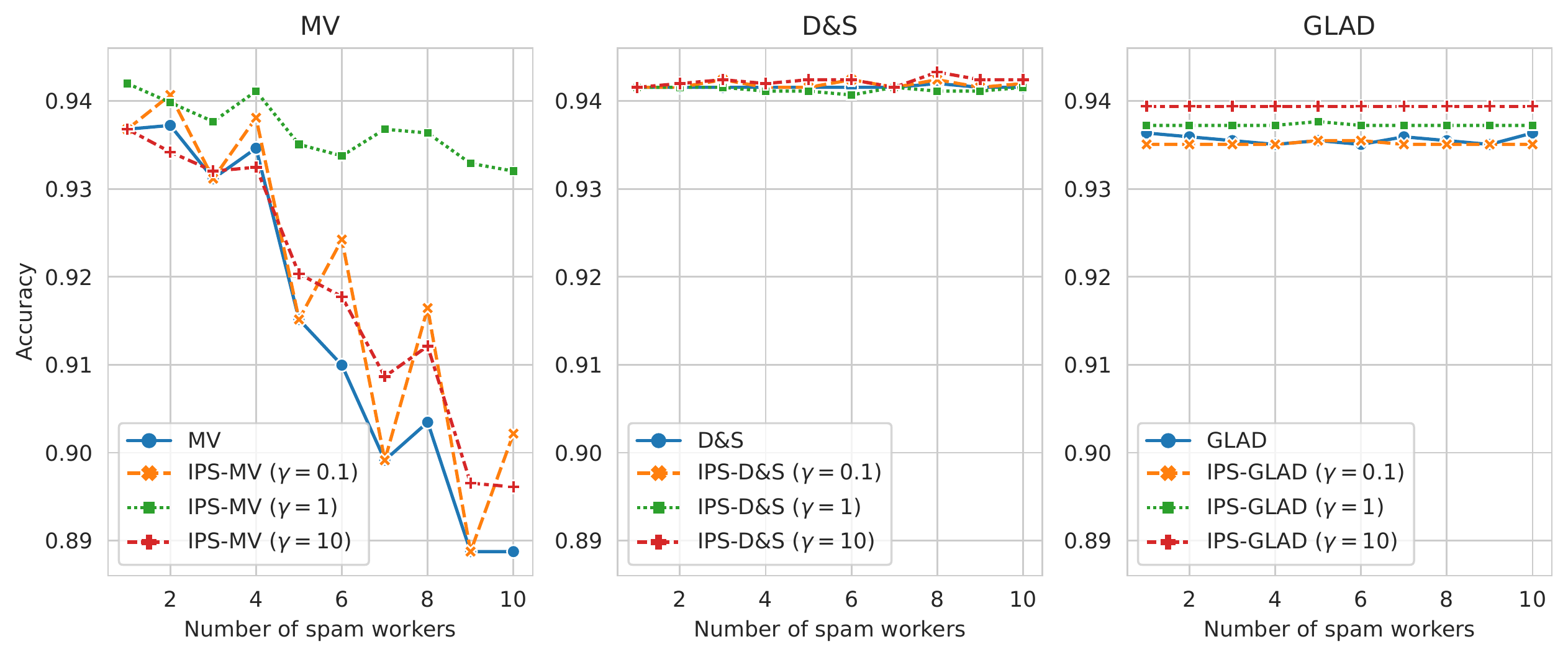}} \\
    \subfloat[WSD\label{fig:spam-WSD}]{\includegraphics[width=0.45\linewidth]{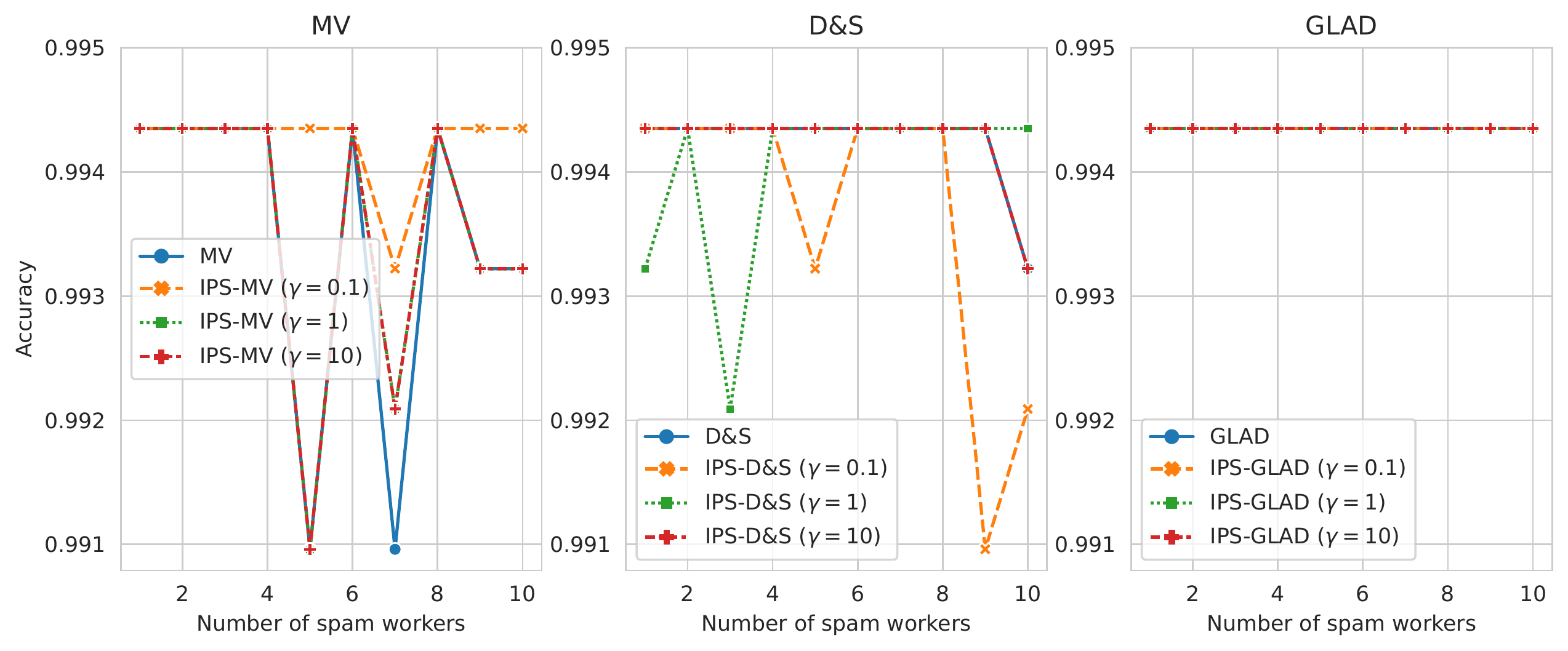}} \quad
    \subfloat[SP\label{fig:spam-SP}]{\includegraphics[width=0.45\linewidth]{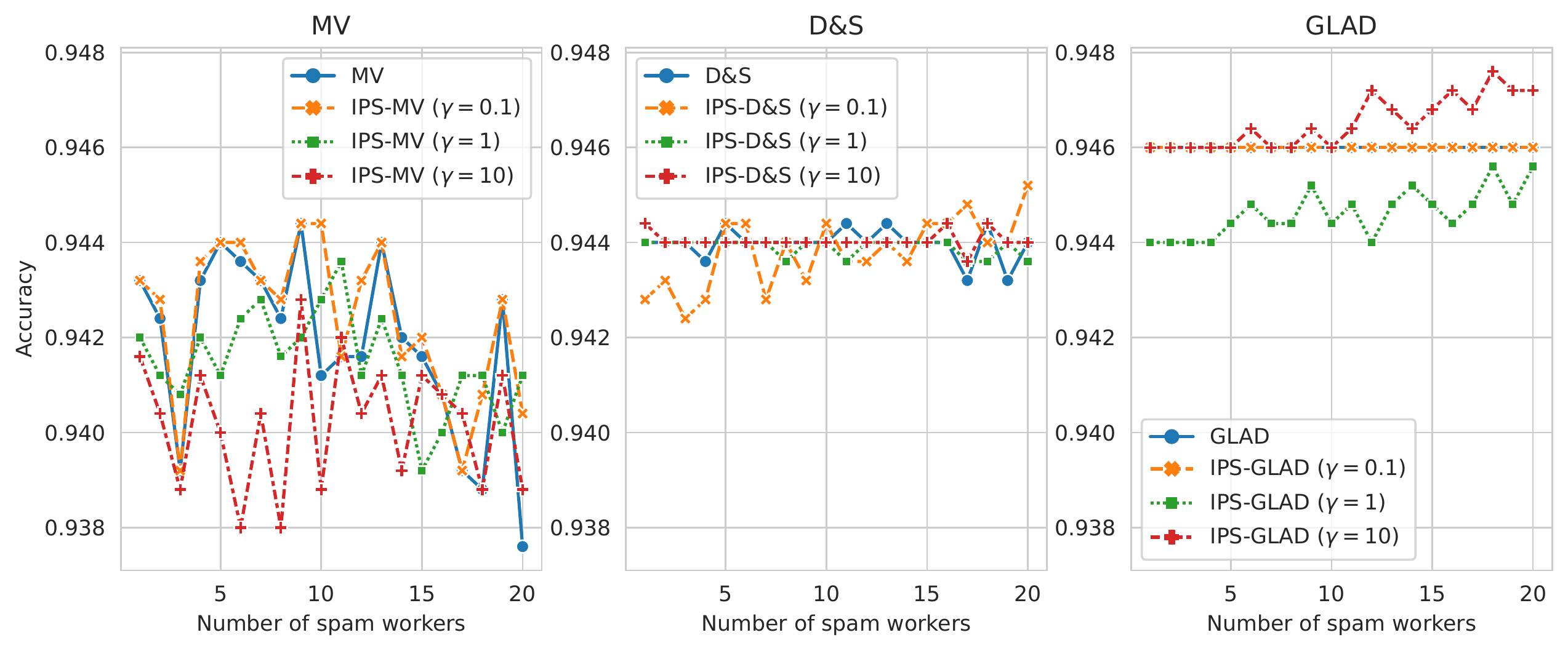}}
    \caption{Robustness against spam workers. }
    \label{fig:spam}
\end{figure*}

\begin{figure*}
    \centering
    \subfloat[RTE\label{fig:colluding-RTE}]{\includegraphics[width=0.45\linewidth]{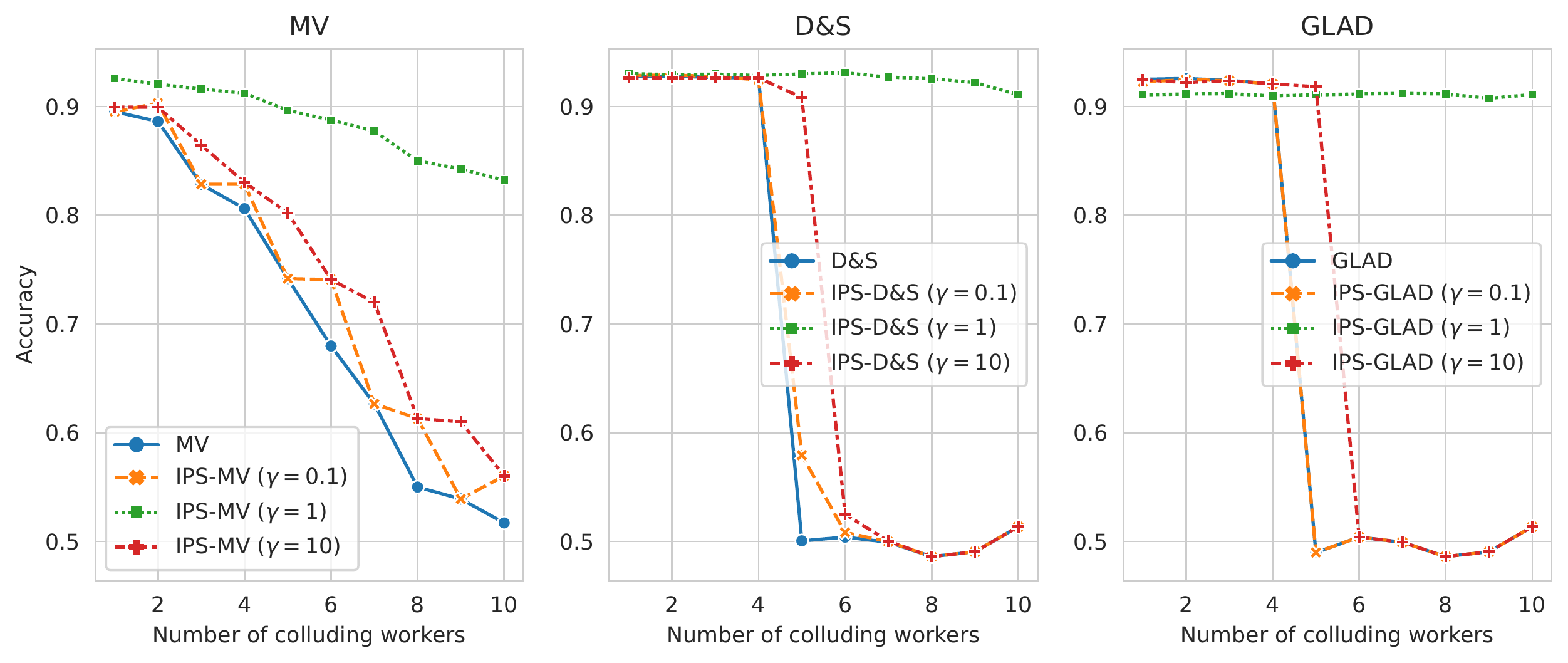}} \quad
    \subfloat[TEMP\label{fig:colluding-TEMP}]{\includegraphics[width=0.45\linewidth]{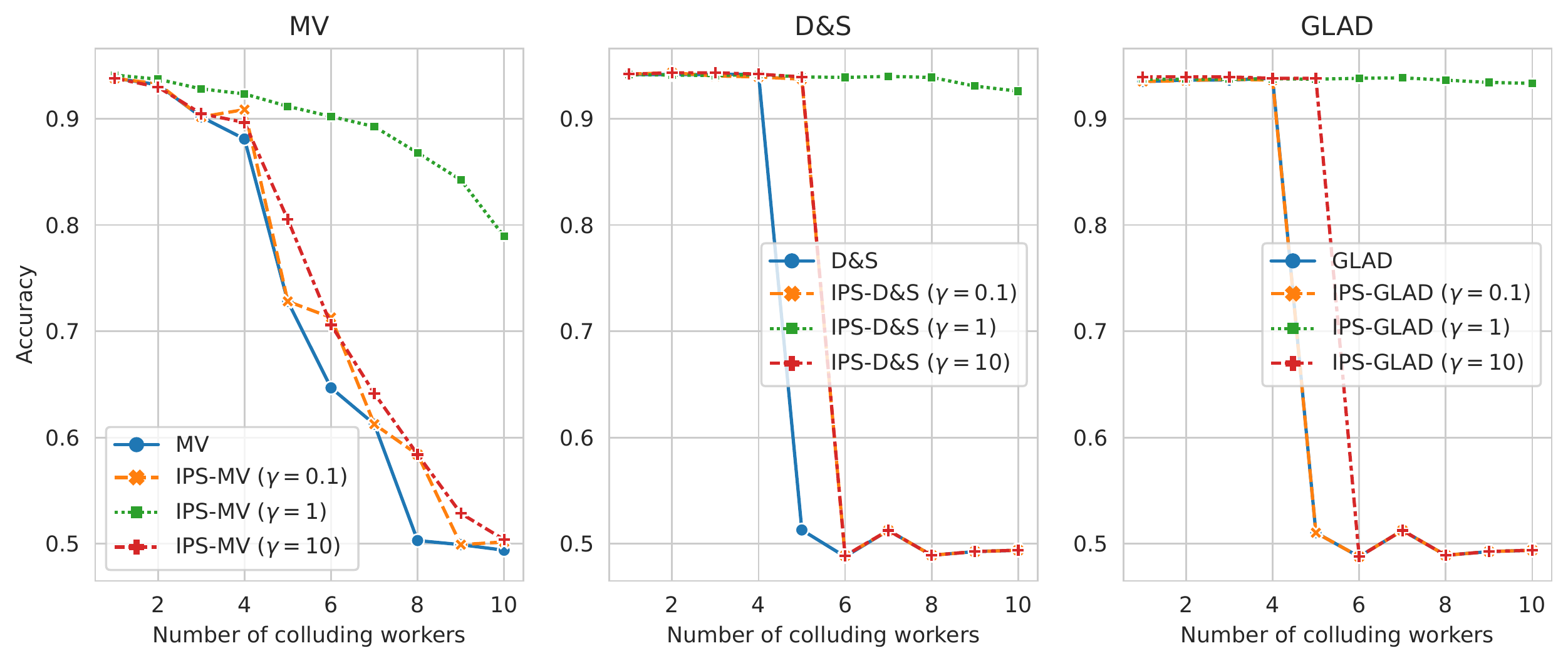}} \\
    \subfloat[WSD\label{fig:colluding-WSD}]{\includegraphics[width=0.45\linewidth]{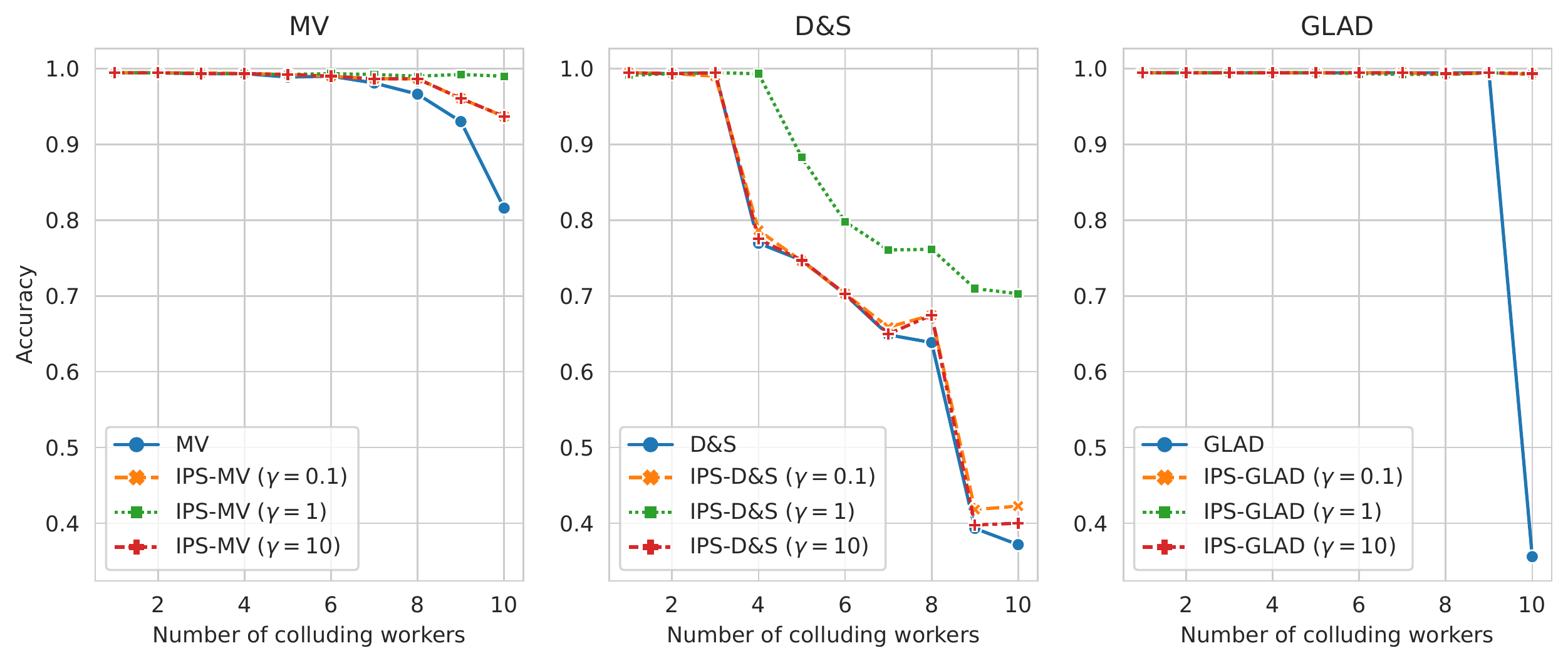}} \quad
    \subfloat[SP\label{fig:colluding-SP}]{\includegraphics[width=0.45\linewidth]{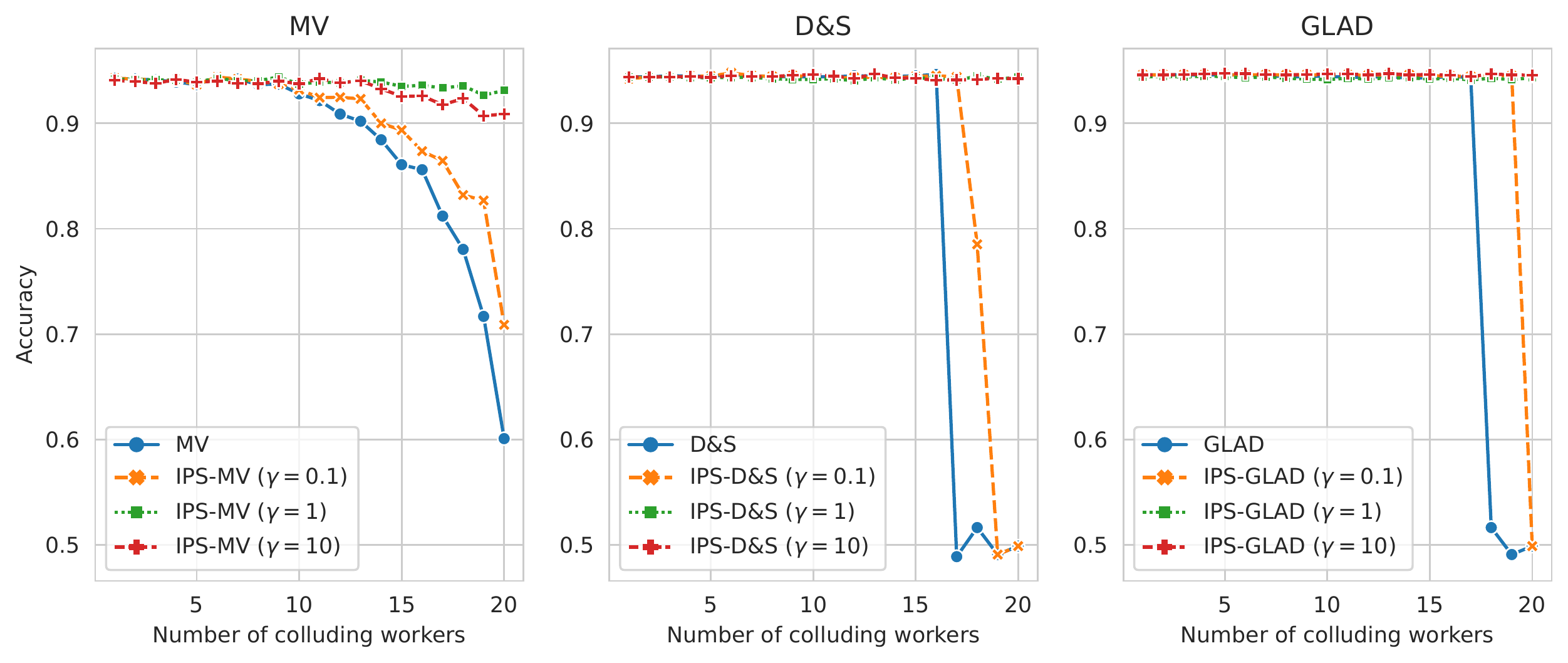}}
    \caption{Robustness against colluding workers. }
    \label{fig:colluding}
\end{figure*}

Figures \ref{fig:spam} and \ref{fig:colluding} show the accuracy when the numbers of added spam workers or colluding workers are varied, respectively.
The accuracy is obtained as the average of five trials of random data generation.
In Figure \ref{fig:spam-RTE} and \ref{fig:spam-TEMP}, with an increase in the number of spam workers, the performance of the simple MV degrades whereas the IPS-MV ($\gamma=1$) remains robust against them.
In Figure \ref{fig:spam-WSD} and \ref{fig:spam-SP}, IPS-MV, and IPS-D\&S show unstable results. This is probably due to the extremely high accuracy in (c) WSD and the large number of original labels in (d) SP.
Figure~\ref{fig:spam} also shows that D\&S, IPS-D\&S, GLAD, and IPS-GLAD are consistently robust against spammers since all of them consider worker's ability.

The robustness against harmful workers is more significant for colluding workers.
Figure~\ref{fig:colluding} shows the accuracy when colluding workers exist. In contrast to the previous experiment, not only MV, but also D\&S and GLAD decreased significantly in accuracy across all the datasets as the number of colluding workers increase. On the other hand, combining the proposed method with MV, D\&S, and GLAD consistently improved performance, especially at $\gamma = 1$.

\section{Conclusion}
We investigated the effect of observation bias, as well as how to deal with such bias in crowdsourcing response aggregation.
By introducing the IPS into the response aggregation methods (majority voting, D\&S and GLAD), we proposed response aggregation methods that eliminate observation bias.
Experiments on synthetic and real data show that the proposed method is effective when a negative correlation exists between the correct answer and the observation rates.
By adding spam and colluding workers to the real datasets, we also demonstrated that the proposed method is robust against such harmful workers.
Since our main focus of this study is to investigate and mitigate the observation bias, we restricted ourselves to the rather classical label aggregation methods.
In the future, we will study more modern and sophisticated aggregation methods.

\section*{Acknowledgment}
This work was supported by JST CREST Grant Number JPMJCR21D1.




%
\balance

\bibliographystyle{IEEEtran}
\bibliography{IEEEabrv,references}

\end{document}